# Carbon Nanotube Initiated Formation of Carbon Nanoscrolls

Zhao Zhang[a], Teng Li[a,b,*]

[a]Department of Mechanical Engineering, University of Maryland, College Park, MD 20742

[b]Maryland NanoCenter, University of Maryland, College Park, MD 20742

**Abstract**

The unique topology and exceptional properties of carbon nanoscrolls (CNSs) have inspired unconventional nano-device concepts, yet the fabrication of CNSs remains rather challenging. Using molecular dynamics simulations, we demonstrate the spontaneous formation of a CNS from graphene on a substrate, initiated by a carbon nanotube (CNT). The rolling of graphene into a CNS is modulated by the CNT size, the carbon-carbon interlayer adhesion, and the graphene-substrate interaction. A phase diagram emerging from the simulations can offer quantitative guideline toward a feasible and robust physical approach to fabricating CNSs.

---

[*] Author to whom correspondence should be addressed. Electronic mail: LiT@umd.edu

A carbon nanoscroll (CNS) is formed by rolling up a graphene sheet into a spiral multilayer structure.[1-3] CNSs are topologically open. For example, the core size of a CNS can vary significantly by relative sliding between adjacent layers.[4,5] By contrast, a multiwall carbon nanotube (MWCNT) consists of several coaxial carbon cylinders and is topologically closed. The open and highly tunable structure of CNSs, combining with the exceptional mechanical and electronic properties inherited from the basal graphene,[6,7] has inspired potential applications of CNSs, such as hydrogen storage medium,[8,9] water and ion channels,[10] nano-oscillators,[11] and nano-actuators.[12] Enthusiasm aside, the realization of these promising applications hinges upon feasible and reliable fabrication of high quality CNSs, which remains as a significant challenge. Here we use molecular dynamics (MD) simulations to demonstrate a simple physical approach to fabricating CNSs via CNT-initiated rolling of graphene on a substrate.

The experimental discovery of CNSs were achieved via a chemical approach, in which graphite is first intercalated using alkali metals, and the resulting exfoliated graphite sheets can curl into scrolls upon sonication.[2] The chemical approach results in scrolls of graphite sheets with undetermined number of layers, which also suffer from the contamination of residual solvent. The surge of interests in graphene in the past several years has enabled the fabrication of graphene monolayers via mechanical exfoliation.[13] Recent experiments show that a $SiO_2$-supported graphene monolayer immersed in isopropyl alcohol (IPA) solution can roll up to form a CNS.[14] The formation of CNSs is highly sensitive to the concentration of IPA solution and the shape of the graphene. In general, the existing chemical approaches to fabricating CNSs suffer from the possible contamination of chemical residue, and also the difficulty in controlling the rolling initiation and rolling direction. So far, no physical approach to fabricating CNSs has been experimentally demonstrated.

Theoretical analysis and MD simulations have been conducted to investigate mechanisms governing the formation of CNSs from graphene monolayer.[4,15,16] Simulations show that a sufficiently large overlap between two edges of a freestanding graphene monolayer can lead to further relative sliding of the overlapped area and eventually forming a CNS.[4] It is also shown that a long and narrow freestanding carbon nanoribbon can spontaneously form a short CNS driven by low temperature (<100K) thermal fluctuation.[16] Recent simulations demonstrate that water nanodroplets could activate the folding of freestanding graphene to form different carbon nanostructures, including CNSs, depending on the size of both graphene and nanodroplets.[15] These simulation demonstrations, however, are rather challenging to be realized in experiment, given the challenge to manipulate freestanding graphene and the difficulty to control water nanodroplets.

Inspired by recent experiments and simulations,[12,14,15] in this letter we use MD simulations to demonstrate an all-dry physical approach to fabricating CNSs, in which the rolling of a substrate-supported graphene monolayer is initiated by a CNT. Figure 1 depicts the simulation model, in which a CNT is placed along the left edge of a flat rectangular graphene monolayer supported by a $SiO_2$ substrate. The CNT-initiated formation of a CNS from the substrate-supported graphene is governed by the interplay among the following energies: the CNT-graphene interaction energy $E_{tg}$, the graphene-graphene interlayer interaction energy $E_{gg}$ (once graphene starts to roll into a CNS), the graphene strain energy $E_g$, and the graphene-substrate interaction energy $E_{gs}$. The non-bonded CNT-graphene interaction and graphene-graphene interlayer interaction can be characterized by vdW force. The weak interaction between a mechanically exfoliated graphene and its substrate can also be characterized by vdW force. Due to the nature of vdW interaction, $E_{tg}$ and $E_{gg}$ minimize when the carbon-carbon (C-C) interlayer distance reaches an equilibrium value, so does $E_{gs}$ when the distance between the graphene and the substrate surface reaches its equilibrium. When the graphene separates from the substrate, curls up to wrap the CNT, and later starts to roll into a CNS, $E_{tg}$ and $E_{gg}$ decrease; on the other hand, $E_{gs}$ increases, so does $E_g$ due to the mechanical deformation of the graphene associated with the wrapping and rolling. Above said, $E_{tg}$ and $E_{gg}$ serve as the driving force, while $E_{gs}$ and $E_g$ represent the resistant force in the CNT-initiated formation of a CNS from a substrate-supported graphene.

In the simulations, the C-C bonds in the CNT and graphene are described by the second generation Brenner potential.[17] The non-bonded CNT-graphene interaction and graphene-graphene interlayer interaction are described by a Lennard-Jones pair potential $V_{cc}(r) = 4\lambda_{cc}\varepsilon_{cc}(\sigma_{cc}^{12}/r^{12} - \sigma_{cc}^{6}/r^{6})$, where $\varepsilon_{cc} = 0.00284\,\text{eV}$, $\sigma_{cc} = 0.34\,\text{nm}$ and $\lambda_{cc}$ is a tuning factor that is used to vary the C-C interaction energy to study its effect on the CNS formation. It has been shown that the effective C-C interaction energy in a CNS can be tuned by an applied dc/ac electric field [12,18]. The non-bonded graphene-SiO$_2$ substrate interaction is described by a Si-C pair potential and an O-C pair potential, both of which take the same form of $V_{cc}(r)$ but with different parameters, that is, $\varepsilon_{SiC} = 0.00213\,\text{eV}$, $\sigma_{SiC} = 0.15\,\text{nm}$, $\varepsilon_{OC} = 0.00499\,\text{eV}$ and $\sigma_{OC} = 0.23\,\text{nm}$, respectively. The tuning factor for the graphene-substrate interaction $\lambda_{CS}$ is taken to be the same for both Si-C and O-C pair potentials. To reduce the computation size, all atoms in the SiO$_2$ substrate are fixed during simulation. This assumption is justified by the weak graphene-substrate interaction and the rigidity of bulk SiO$_2$. The graphene used in the simulations is 50 nm long and 4 nm wide, with the right edge constrained on the substrate by a linear spring. CNTs with length of 4 nm but various diameters are used to study the effect of CNT size on the CNS formation. The MD simulations are carried out using LAMMPS[19] with NVT ensemble at temperature 300K and with time step 1 fs.

Depending on the CNT size, C-C interaction strength and C-SiO$_2$ interaction strength, the CNT-graphene-substrate system shown in Fig. 1 evolves in three different modes. Figure 2 illustrates the time sequential snapshots of each mode of evolution and the corresponding variation in the total potential energy of the system. In the three cases shown in Fig. 2, $\lambda_{CC} = 1$ and $\lambda_{CS} = 1$. CNTs of various diameters (i.e., (10,10), (12,12) and (18,18)) are used, respectively.

The graphene strain energy $E_g$ due to wrapping a CNT is roughly inversely proportional to the square of the CNT diameter. If the CNT diameter is too small, the significant increase of $E_g$ and the corresponding increase of $E_{gs}$ due to graphene-substrate separation can overbalance the decrease of $E_{tg}$ due to graphene wrapping the CNT. As a result, instead of wrapping the CNT, the graphene remains flat on the substrate, while the CNT glides on the graphene driven by the thermal fluctuation, as shown in Fig. 2b-e. The translational motion of the CNT on the graphene is expected to cause negligible variation to the total potential energy of the system except the thermal fluctuation, which is evident in Fig. 2a.

If a CNT of intermediate diameter is used, the increase of $E_g$ due to graphene bending and $E_{gs}$ due to graphene-substrate separation can be outweighed by the corresponding decrease of $E_{tg}$. Consequently, graphene can separate from the substrate under thermal fluctuation and start to wrap around the CNT (Fig. 2g). The total potential energy continues decreasing until the whole surface of the CNT is nearly wrapped by the graphene (Fig. 2h). Further rolling of the graphene is hindered by a local energy barrier due to the step formed by the left edge of the graphene adhering to the CNT. If the CNT radius is not sufficiently large, the energy barrier due to the graphene edge step can be too high and thus prevent the further rolling of graphene. This is analogous to a roller moving toward a speed bump of a fixed height. If the roller is too thin, instead of passing the bump, it can be bounced up in the air. As shown in Fig. 2h-i, the graphene-wrapped CNT rolls toward the graphene edge step, and is then bounced upward. The kinetic energy of the graphene-wrapped CNT leads to further separation of a short segment of graphene from the substrate, and then the graphene-substrate interaction pulls the separated graphene re-adhered back to the substrate. Such two processes repeat several times and the graphene segment eventually re-adheres back to the substrate after the excess translational kinetic energy is dissipated. The CNT remains wrapped by the graphene (Fig. 2j).

If the diameter of the CNT is sufficiently large, the translational kinetic energy of the graphene-wrapped CNT can overcome the fixed energy barrier due to the graphene edge step. As a result, CNT-initiated rolling of the graphene continues and then an overlap between the left edge and the flat portion of the graphene forms. Such an overlap leads to the decrease of $E_{gg}$, which drives further rolling of graphene into a CNS (Fig. 2l-o, and also the supplemental materials). As shown in Fig. 2k, the continuous rolling of the graphene results in further decrease of the overall potential energy, which is more substantial than that due to the graphene wrapping the CNT. The resulting CNS is energetically stable against thermal perturbations at 300 K.

When the CNT size and C-SiO$_2$ interaction strength are fixed, the evolution of the CNT-graphene-substrate system can be modulated by the C-C interaction strength. Figure 3a defines a phase diagram of the evolution of the CNT-graphene-substrate system in the space of C-C interaction strength and CNT size, for a given C-SiO$_2$ interaction strength (i.e., $\lambda_{CS} = 1$). The same three modes of evolution as described above are observed. For a given CNT size, the mode of evolution changes from CNT gliding to graphene wrapping and then to graphene forming a CNS, as the C-C interaction becomes stronger. For a given C-C interaction strength, the similar change of the mode of evolution is shown as the CNT size increases. Emerging from the simulations are a boundary between mode I and mode II and that between mode II and mode III, the latter of which can serve as a guidance for controlling CNS formation by varying C-C interaction and selecting CNT size. Figure 3b plots the case of $\lambda_{cs} = 4$. When the C-SiO$_2$ interaction strength increases, a stronger C-C interaction is needed for the CNS formation, for a given CNT size; similarly, a CNT with larger diameter is needed to initiate the CNS formation, for a given C-C interaction strength. For the case of $\lambda_{cs} = 4$, there exists a critical value of $\lambda_{cc}$, below which graphene rolling into a CNS cannot be initiated by a CNT of any given size.

In summary, we demonstrate the CNT-initiated formation of a CNS from substrate-supported graphene, using MD simulations. The CNT is shown to help overcome the energy barrier to form an overlap in graphene. Once the overlap is formed, the graphene can spontaneously roll up into a CNS. The successful formation of a CNS depends on the CNT diameter, the C-C interaction strength and the graphene-substrate interaction strength. The phase diagram obtained from this study elucidates critical parameters governing the formation of CNSs from graphene. With the ever maturing fabrication of high quality CNTs and large area graphene on substrates, and the nanopatterning technique to position these building blocks at high precision, the CNT-initiated formation of CNSs holds great potential leading to a feasible, all-dry, physical fabrication technique of high quality CNSs. The resulting CNS nanostructures hold potential to enable unconventional nanoscale electromechanical devices.[12]

**Acknowledgement:** This work is supported by a UMD GRB summer research award and NSF Grant No. 0928278. Z.Z. also thanks the support of A. J. Clark Fellowship and UMD Clark School Future Faculty Program.

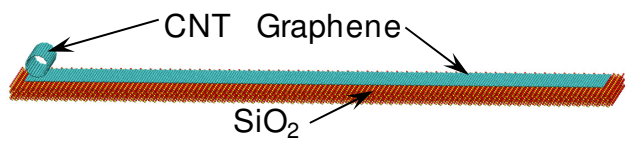

FIG. 1. (Color online) The MD simulation model. A graphene is supported by a $SiO_2$ substrate, with a CNT placed along the left edge of the graphene.

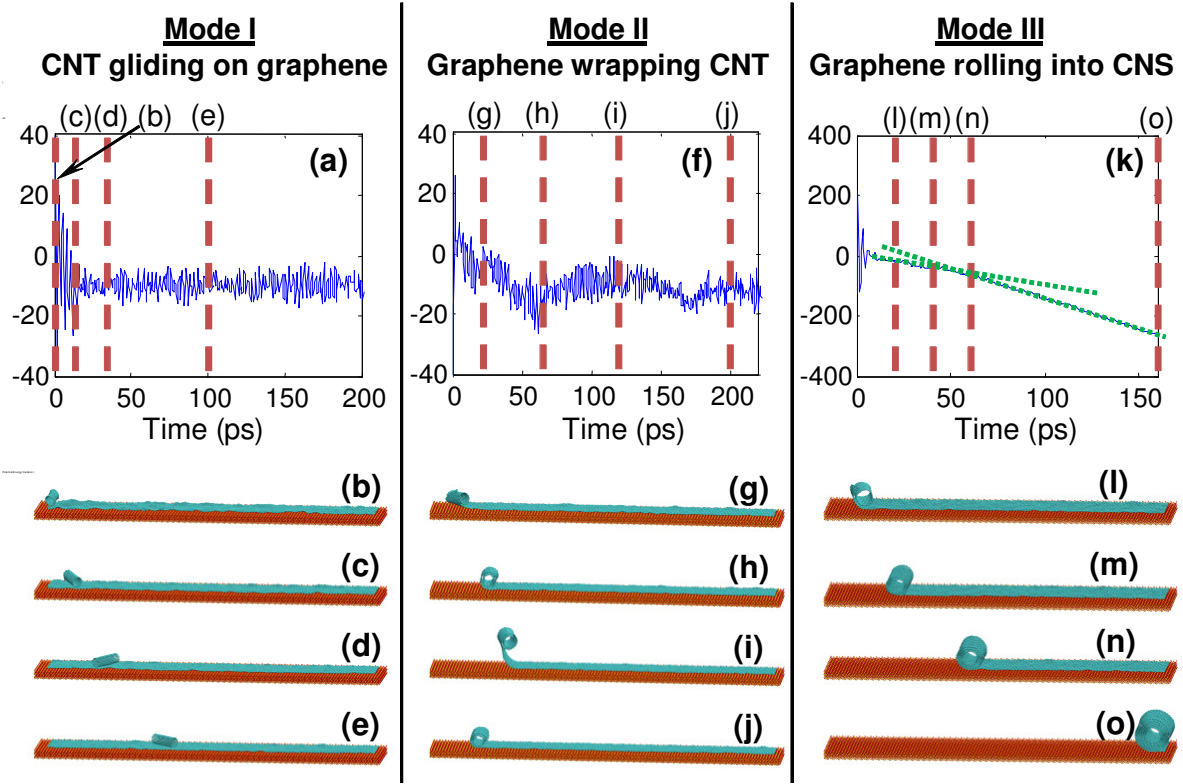

FIG. 2. (Color online) Three modes of evolution of the CNT-graphene-substrate system. (a, f, k) plot the variation in the total potential energy of the system as a function of simulation time for each mode, respectively; (b-e): Snapshots of a (10, 10) CNT gliding on the substrate-supported graphene at 0 ps, 15 ps, 40 ps and 100 ps, respectively; (g-j): Snapshots of the graphene wrapping a (12, 12) CNT at 20 ps, 65 ps, 120 ps and 200 ps, respectively; (l-o): Snapshots of the graphene rolling into a CNS, initiated by a (18,18) CNT, at 20 ps, 45 ps, 60 ps and 160 ps, respectively. The two dotted fitting curves in (k) show that the graphene further rolling into a CNS (from (n) to (o)) leads to more substantial decrease of potential energy than that due to graphene wrapping CNT (from (l) to (m)). In all three cases shown here, $\lambda_{CC} = 1$ and $\lambda_{CS} = 1$. (enhanced online).

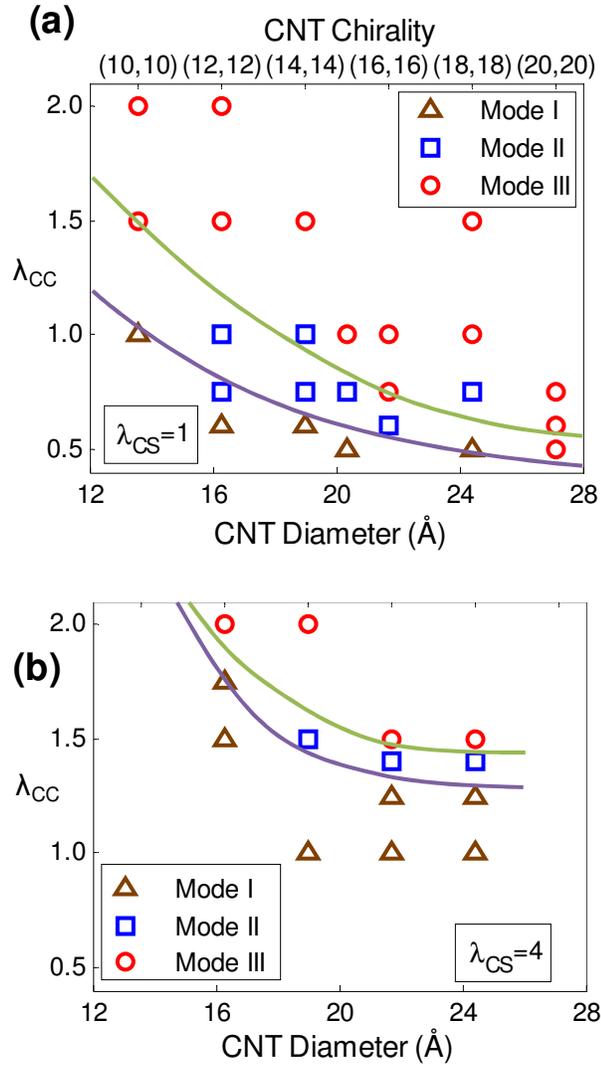

FIG. 3. (Color online) Phase diagrams of the evolution of the CNT-graphene-substrate system in the space of C-C interaction strength and CNT size, for a given C-SiO$_2$ interaction strength. Here, (a) $\lambda_{CS} = 1$, (b) $\lambda_{CS} = 4$.